\documentclass[aps,prd,twocolumn,preprintnumbers,groupedaddress,nofootinbib]{revtex4}
\usepackage{amsmath}
\usepackage{amsfonts}
\usepackage{amssymb}
\usepackage{mathrsfs}
\usepackage{graphicx}
\usepackage{multirow}
\usepackage{color}
\usepackage{hyperref}
          \newcommand{\etal}{{\it et al. }}
          
           \newcommand{\GeV}{{\mathrm {GeV}}}

   \newcommand{\MeV}{{\mathrm {MeV}}}
     \newcommand{\eV}{{\mathrm {eV}}}
    \newcommand{\fb}{{\mathrm {fb}}}
\newcommand{\dd}{\mathrm{d}}   \newcommand{\TeV}{{\mathrm {TeV}}}
\newcommand{\mb}[1]{\boldsymbol{#1}}
 \newcommand{\rep}[1]{\mathbf{#1}}
\newcommand{\conjrep}[1]{\overline{\mathbf{#1}}}

\begin{document}
\title{R-symmetric High Scale Supersymmetry\footnote{Incorporates aspects of preprint \cite{Unwin:2011ag}, with substantial new material.}
}
\author{James Unwin}
\email{unwin@maths.ox.ac.uk}
\affiliation{Mathematical Institute, University of Oxford,
24-29 St Giles', Oxford, OX1 3LB, UK\\
Rudolf Peierls Centre for Theoretical Physics, University of Oxford,
1 Keble Road, Oxford, OX1 3NP, UK}

\begin{abstract}

Introducing an $R$-symmetry to models of high scale supersymmetry (SUSY) can have interesting consequences, and we focus on two aspects.  If Majorana masses are forbidden by an $R$-symmetry and the main source of electroweak gaugino masses are Dirac terms, then the Higgs quartic coupling vanishing at the SUSY scale and the Higgs boson mass will be near 125 GeV. Moreover, using an $R$-symmetry, models with only one Higgs doublet in the UV can be constructed and we argue that, since we desire only a single Higgs at the weak scale, this scenario is more aesthetic than existing models. We subsequently present a model which draws on both of these features. We comment on neutrino masses and dark matter in these scenarios and discuss how the models presented can be discerned from alternative constructions with high scale SUSY, including Split SUSY.

\end{abstract}

\maketitle

\section{Introduction}

\vspace{-2mm}
Supersymmetry (SUSY) is a compelling theoretical proposal, being the unique extension of space-time geometry beyond the Poincar\'e group. The maximal amount of SUSY at low energy which is compatible with the Standard Model (SM) is found for an $\mathcal{N}=1$ matter sector, with an $\mathcal{N}=2$ gauge sector and this scenario can be realised by supplementing  the familiar MSSM spectrum with adjoint chiral superfields \cite{Fox:2002bu,Dirac,Abel:2011dc,Benakli:2010gi}. The scale of the SUSY breaking, however, is an undetermined parameter and it has previously been argued \cite{Dine} that very high scale SUSY breaking, approaching the fundamental Planck scale, is favoured in string constructions. If this were the case, then no apparent signs of SUSY would remain in the low energy theory.

 Split SUSY \cite{split}, Environmentally Selected SUSY Standard Models (E-SSM) as proposed by Hall \& Nomura, \etal \cite{Hall:2009nd,Elor:2009jp}, and related models \cite{Other}, are built on the premise that the hierarchy problem can be explained through environmental selection on the scale of electroweak symmetry breaking \cite{hier}.  Environmental selection arguments occur quite naturally in the context of the string theory landscape and have previously been used to suggest resolutions to a variety of problems, most prominently the cosmological constant \cite{Weinberg:1987dv}. 
 It is further argued that whilst SUSY is no longer needed to solve the hierarchy problem, it should be present at high scales since it is crucial in any physical realisation of  string theory.  In these models the low energy spectra contain only the SM states, including a single Higgs boson, and (possibly) a selection of superpartners which can  provide the dark matter (DM) relic density (e.g.~winos). The other states in the theory acquire masses near the SUSY breaking scale $\widetilde{m}$. Split  SUSY and E-SSM are closely related, the frameworks differ mainly in their low energy spectra, Split SUSY generically has TeV Higgsinos, due to a small $\mu$ term, in order to improve gauge coupling unification.

Although, na\"ively, high scale SUSY breaking leaves no trace of its existence at low energy, it has been argued that predictions of the Higgs boson mass can be made and these have been previously calculated in e.g.~\cite{Giudice:2011cg}  
\begin{equation}
\begin{aligned}
m_H^{\mathrm{Split\, SUSY}}\simeq&\, 140\pm15\, \GeV~, \\
 m_H^{\mathrm{E-SSM}}\simeq&\, 133\pm10\, \GeV~.
\label{1}
\end{aligned}
\end{equation}
These values are derived by matching the Higgs quartic coupling $\lambda_H$ with the  SUSY boundary condition at the SUSY breaking scale
{\small{\begin{equation}
\lambda_H=\frac{(g^2+g^{\prime}{}^2)}{8}~\cos^22\beta~,
\label{2}
\end{equation}
}}
and using renormalisation group methods to scale the couplings from the SUSY breaking scale to the weak scale. Notably the a major source of uncertainty in these Higgs mass calculations is the unknown value of $\tan\beta$ and if the $\tan\beta$ dependence is removed then the Higgs mass may be predicted to within a few GeV.

In this paper we explore the possible consequences of introducing an $R$-symmetry to models of high scale SUSY. In Sect.~\ref{sec2} we propose a new class of high scale SUSY models with an $R$-symmetry, based on the SUSY One Higgs Doublet Model (SOHDM) \cite{Davies:2011mp}, and we refer to the resulting construction as E-SOHDM. We argue that this is a more appealing setting in which to realise a single scalar Higgs in the IR theory, as the UV theory contains from the outset only one Higgs doublet (i.e.~a doublet with couplings to SM fermions and obtaining a VEV in the IR).  A second scenario which can naturally occur in models with an $R$-symmetry is that the gauginos can be nearly pure Dirac states  \cite{Rsym} and in Sect.~\ref{sec3} we demonstrate that this automatically results in the correct Higgs mass if the SUSY scale is around $10^{10\pm1}$ GeV. Moreover, both of these frameworks remove the $\tan\beta$ dependence inherent to the SUSY boundary condition and thus result in much sharper predictions of the Higgs boson mass compared to Split SUSY or E-SSM. In Sect.~\ref{sec4} we present a model which incorporates both of these ideas and in Sect.~\ref{sec5} we discuss how different proposals with high scale SUSY breaking might be distinguished.

%%%%%%%%%%%%%%%%%%%%%%%%
\section{E-SOHDM}
\label{sec2}

Following Split  SUSY and E-SSM, we assume that the hierarchy problem is explained through fine-tuning due to environmental selection and further that anthropic requirements also determine the DM relic density. 
In analogy with existing models we seek a low energy spectrum which features only the SM states, including a single Higgs boson, and TeV scale gauginos provide the DM. Specifically, a neutral wino  $\widetilde{W}^0$ LSP provides a favourable weakly interaction massive particle (WIMP) candidate for DM. The scenario which is most naturally realised in this model is a TeV scale neutral wino LSP which is nearly degenerate with the charged winos. In this case, since $m_Z\ll m_{\widetilde{W}^0}$, the wino annihilation cross section is Sommerfeld enhanced and this causes a reduction in the wino thermal relic abundance. In order to reproduce the observed DM relic density, at $2 \sigma$, the wino mass must lie in the range \cite{Hisano:2006nn}:
\begin{equation}
2.5\,\TeV\lesssim m_{\widetilde{W}^0} \lesssim 3.0\, \TeV~.
\label{DM}
\end{equation}
Note that this result includes effects due to coannihilation.
Whilst it is not possible to observe 2.5 TeV winos at current direct detection experiments, current indirect detection projects and next generation direct detection experiments may be able to test this prediction. The experimental signals are discussed in \cite{Elor:2009jp}.

Since we wish to have one Higgs boson in the low energy spectrum we shall insist that only a single Higgs field is present in the model. Normally, two Higgs doublets are required in (minimal) SUSY theories in order to give masses to the quarks and leptons, and to ensure anomaly cancellation. However, a single Higgs doublet, the scalar component of  $\mb{H_u}$, can provide masses to all of the SM fermions \cite{Davies:2011mp,OHD}. Following  \cite{Davies:2011mp}, the chiral superfield $\mb{H_d}$ is included to cancel anomalies, although it does not obtain a VEV and symmetries forbid Yukawa couplings involving $\mb{H_d}$. The field $\mb{H_d}$ can no longer be considered a Higgs and consequently is relabelled $\mb{\eta}$. The field $\mb{H_u}$, being the only true Higgs field, is labelled $\mb{H}$. The field content and charges of the chiral superfields are summarised in Tab.~\ref{Tab1}. The chiral superfield $\mb{X}$ is a spurion field which parameterises the SUSY breaking. The model has an anomaly free approximate U(1) $R$-symmetry and matter parity in order to stabilise the LSP. Note that the symmetries forbid Majorana gaugino mass terms and trilinear $\mathcal{A}$-terms, but allow the $\mu$ term
\begin{equation}
\begin{aligned}
\mathcal{L}_{\mu}=
\int\mathrm{d}^4\theta & \frac{\mb{X^{\dagger}}}{M_*}
 \lambda_\mu\mb{H\eta}~.
\label{higgs}
\end{aligned}
\end{equation}
This leads to an effective $\mu$ of order $\frac{F_X}{M_*}\sim\widetilde{m}$, where  $F_X$ is the $F$-term SUSY breaking expectation value of $\mb{X}$.  However, since we do not require the Higgsinos to lie near the weak scale, there is no $\mu$-problem. Note that the scale of $\mu$ is the main difference between the class of models presented here and Split SUSY. The size of $\mu$  leads to deviations in the Higgs mass between the two frameworks. Moreover, the lightest neutralino in models of Split SUSY is an unknown mixture of the neutral Higgsinos and gauginos \cite{Wang:2005kf}. Since here we have $\mu\sim\widetilde{m}$, the lightest neutralino is almost completely wino and (assuming that this state is responsible for the DM density) this results in a much sharper prediction of the DM mass. Some aspects of Split SUSY models with large $\mu$ were previously studied in \cite{Cheung:2005ba}.

%%%%%%%%%%%%%%%%%%%%%%%%%%%%%%%%
%
%
\begin{table}[t]
\begin{center}
\def\str{\vrule height12pt width0pt depth7pt}
\begin{tabular}{| c | l | c | c | }
    \hline\str
    ~Field ~&~ Gauge rep. ~&~ U(1)${}_R$ ~&~ $(-1)^{3(B-L)}$~ \\
    \hline\str
    ~~$\mb{Q}$ & ~~$\,(\rep{3},\rep{2})_{1/6}$ & $1$ & $-$\\
    \hline\str
    ~~$\mb{U^c}$ & ~~$\,(\conjrep{3},\rep{1})_{-2/3}$ & $1$ & $-$ \\
    \hline\str
    ~~$\mb{D^c}$ & ~~$\,(\conjrep{3},\rep{1})_{1/3}$ & $1$ & $-$\\
    \hline\str
    ~~$\mb{L}$ & ~~$\,(\rep{1},\rep{2})_{-1/2}$ & $1$ & $-$ \\
    \hline\str
    ~~$\mb{E^c}$ & ~~$\,(\rep{1},\rep{1})_1$ & $1$ & $-$\\
    \hline\str
    ~~$\mb{H}$ & ~~$\,(\rep{1},\rep{2})_{1/2}$ & $0$ & $+$ \\
    \hline\str
    ~~$\mb{\eta}$ & ~~$\,(\rep{1},\rep{2})_{-1/2}$ & $2$ & $+$\\
      \hline\str
    ~~$\mb{X}$ & ~~$\,(\rep{1},\rep{1})_0$ & $2$ & $+$\\
    \hline
\end{tabular}
\caption{Spectrum of chiral superfields \cite{Davies:2011mp}.}
\label{Tab1}
\end{center}
\end{table}
%
%
%%%%%%%%%%%%%%%%%%%%%%%%%%%%%%%%%%

The  U(1) $R$-symmetry forbids Majorana gaugino mass terms. However, suitable gaugino masses can be generated via the model-independent contribution from $R$-symmetry breaking due to supergravity \cite{Giudice:1998xp,Randall:1998uk} 
\begin{equation}
M_i\simeq\frac{g^2_i}{16\pi^2}b_0^im_{3/2}~,
\end{equation}
where $b_0^i$ are the $\beta$ function coefficients of the gauge couplings. Note that as the gauginos are the only non-SM states introduced below $\widetilde{m}$ the $\beta$ function coefficients are different to the MSSM and closer to the SM:
$b_0^i=\left(-\frac{41}{10},\,\frac{11}{6},\,5\right)$.
Consequently, below $\widetilde{m}$ the running of the coupling constants is comparable to the SM and approximate gauge coupling unification occurs above the scale $\widetilde{m}$, around $10^{17\pm1}$ GeV, with similar precision to SM unification  \cite{Cheung:2005ba}. Furthermore, from the $\beta$ function coefficients we can calculate the gaugino masses
\begin{equation}
\begin{aligned}
|M_1|&
\simeq 3\,\TeV~\left(\frac{m_{3/2}}{550~\TeV}\right),\\
|M_2|&
\simeq 2.75\,\TeV~\left(\frac{m_{3/2}}{550~\TeV}\right),\\
|M_3|&
\simeq 24\,\TeV~\left(\frac{m_{3/2}}{550~\TeV}\right).
\label{w}
\end{aligned}
\end{equation}
For high scale SUSY breaking $M_{1,2}\gg m_Z$, thus the values of $|M_i|$ correspond very well to the masses of the physical gauginos and the wino annihilation cross section is Sommerfeld enhanced, as anticipated. Also, the neutral wino is the LSP and is nearly degenerate with the charged winos, a mass splitting of $\sim165$ MeV is generated by electroweak  corrections \cite{Pierce:1996zz}. We observe that the neutral wino has the correct mass to generate the observed DM relic density if the gravitino mass is $m_{3/2}\sim$ 500 - 600 TeV.
The gravitino mass is given by
\begin{equation}
 m_{3/2}=\frac{F_X}{\sqrt{3}M_{\mathrm{Pl}}}~,
\label{3/2}
 \end{equation}
 where $M_{\mathrm{Pl}}\simeq2.4\times10^{18}$ GeV is the reduced Planck mass. Thus to obtain a suitable gravitino mass we require that the SUSY breaking scale is
\begin{equation}
\sqrt{F_X}\simeq 2\times10^{12}~\GeV~.
\label{FX}
 \end{equation}
This scale is much higher than in models of weak scale SUSY and we shall assume that $F_X$ takes this value in the remainder of the paper, unless stated otherwise.

The SM fermion masses arise from the following Yukawa terms \cite{Davies:2011mp,OHD}
\begin{equation}
\begin{aligned}
\mathcal{L}_{\mathrm{Y}}=&\int\mathrm{d}^2\theta \lambda_U \mb{HQU^c}\\
&+\int\mathrm{d}^4\theta
\frac{\mb{X^{\dagger}}\mb{H^{\dagger}}}{M_*^2}\left( \lambda_D \mb{QD^c}+ \lambda_E \mb{QE^c}\right)~.
\end{aligned}
\end{equation}
All of the quarks and leptons acquire their masses from the VEV of $H$, the scalar component of $\mb{H}$,
\begin{equation}
\langle H\rangle\simeq v/\sqrt{2}\simeq 174\,\GeV~.
\end{equation}
To obtain the observed SM fermion masses, the following tree-level relationship must be satisfied:
\begin{equation}
\frac{\lambda_b F_X}{M_*^2}\times174\GeV \simeq 5\,\GeV~.
\label{con}
\end{equation}

Moreover, the scale $M_*$ naturally provides suitable neutrino masses via the dimension 5 Weinberg operator
\begin{equation}
\begin{aligned}
\mathcal{L}_{\nu}=&\int\mathrm{d}^4\theta 
\frac{\mb{X^{\dagger}}}{M_*^3} \mb{H^2L^2}~.
\end{aligned}
\end{equation}
This term leads to neutrino mass scale of the order
\begin{equation}
M_{\nu}\sim \frac{F_X v^2}{2M_*^3}~.
\end{equation}
To accommodate the observed neutrino scale we require that
$ 0.01\, \eV \lesssim M_\nu \lesssim 1\, \eV$ \cite{neutrinos}.
Comparing with eq.~(\ref{con}), this implies that
\begin{equation}
M_\nu\sim\frac{\TeV}{\lambda_b M_*}
\label{XX}
\end{equation}
and, assuming natural values for the coupling $\lambda_b\sim1$, we observe that phenomenological acceptable neutrino masses can be generated for $M_*\sim10^{13\pm1}$ GeV.
From an anthropic perspective, neutrino masses much higher then this greatly suppress structure formation due to free streaming  \cite{Tegmark:2003ug}, which presents a catastrophic boundary in the landscape and an anthropic constraint on the magnitude of the scale $M_* \gtrsim 10^{12}~ \GeV$.

Comparing the condition $M_*\sim10^{13\pm1}$ GeV with eq.~(\ref{FX}), we find that SUSY scale should be $\widetilde{m}\sim10^{11\pm1}$ GeV. Moreover, we argue that this is the natural scale for SUSY breaking to occur, given that it is related to the fundamental Planck scale. In the context of string theory the compactification scale is related to the fundamental UV scale $M_*$ by  \cite{ArkaniHamed:1998rs} (in the case without warping)
\begin{equation}
M_*^{D+2}\sim\frac{M_{\mathrm{Pl}}^2}{\mathcal{V}}~.
\end{equation}
where  $\mathcal{V}$ is the $D$-dimensional compactification volume.
With a large compactification volume we can obtain a suitable  $M_*\ll M_{\mathrm{Pl}}$.
Moreover, it has been argued \cite{Dine} that, with reasonable assumptions, having the SUSY breaking scale close to the UV cutoff is favourable and it is natural that the scales $M_*$ and $\widetilde{m}$ are comparable.

Next we consider the Higgs mass; the low energy Higgs potential is given by
\begin{equation}
V_H=
-\frac{m_H^2}{2}|H|^2+\frac{\lambda_H}{4}|H|^4~,
\end{equation}
where $m_H$ the physical Higgs boson mass and may be expressed as follows
\begin{equation}
m_H^2=\frac{\lambda_Hv^2}{2}~.
\end{equation}
The Higgs quartic coupling is determined by the SUSY boundary condition and the Higgs VEV is fixed by environmental selection on the size of weak scale which fine-tunes the Higgs soft mass $\widetilde{m}_H$ and the scale $\mu$
\begin{equation}
 \frac{v}{\sqrt{2}}=2\sqrt{\frac{\widetilde{m}_H^2-|\mu|^2}{g^2+g^{\prime}{}^2}}~.
\end{equation}

The value of the Higgs mass may be calculated by noting that  the quartic Higgs coupling is fixed by the SUSY boundary condition at the scale $\widetilde{m}$
\begin{equation}
\lambda_H(\widetilde{m})\simeq\frac{g^2(\widetilde{m})+g^{\prime}{}^2(\widetilde{m})}{8}(1+\delta(\widetilde{m}))~,
\label{ssbc}
\end{equation}
where the quantity $\delta$ accounts for threshold corrections at the scale $\widetilde{m}$. Convergence of the IR flow makes the Higgs mass relatively insensitive to $\delta$ and numerical studies \cite{Hall:2009nd,Elor:2009jp} suggest that UV threshold corrections $\delta$ which effect the value of $\lambda_H(\widetilde{m})$ feed into the Higgs mass via
\begin{equation}
\delta m_H\sim 0.1~\GeV\left(\frac{\delta}{0.01}\right)~.
\end{equation}
Renormalisation group scaling can be used to run all of the couplings from $\widetilde{m}$ to the weak scale in order to determine the physical Higgs mass. The analyses of  \cite{Hall:2009nd,Elor:2009jp} included one loop weak scale threshold corrections (including TeV winos in \cite{Elor:2009jp}), and two and three loop QCD effects. The main sources of uncertainty for the Higgs mass comes from the top mass $m_t$ and QCD coupling $\alpha_S(M_Z)$.  The current experimental values for these quantities are \cite{Tevatron}:
\begin{equation}
\begin{aligned}
m_t &=173.1\pm0.9 \,\GeV,\\
\alpha_S(M_Z) &=0.1184\pm0.0007~.
\label{mt}
\end{aligned}
\end{equation}
By construction, the spectrum below $\widetilde{m}$ is relatively unchanged from certain formulations of the E-SSM. Since the couplings of the gluino and bino to the SM Higgs boson is only through loops involving (heavy) sfermions, the Higgs mass calculation is analogous to the E-SSM with TeV winos studied in \cite{Elor:2009jp} in the limit $\tan\beta\rightarrow\infty$. Here we recapitulate the relevant result of \cite{Elor:2009jp} with updated errors:
\begin{equation}
\begin{aligned}
m_H\simeq&~141\,\GeV \\
&+1.3\, \GeV \left(\frac{m_t-173.1 \,\GeV}{0.9\, \GeV}\right)\\
&-0.35\, \GeV \left(\frac{\alpha_s(M_Z)-0.1176}{0.0007}\right)\\
&+0.14\, \GeV \log_{10}\left(\frac{\widetilde{m}}{10^{11} \,\GeV}\right)~.
\end{aligned}
\end{equation}
\begin{table}[t]
\begin{center}
\def\str{\vrule height10pt width0pt depth7pt}
\begin{tabular}{| c | l | c | c | }
    \hline\str
    ~Field ~&~ Gauge rep. ~&~ U(1)${}_R$ ~&~ $(-1)^{3(B-L)}$~ \\
    \hline\str
    ~~$\mb{T}$ & ~~$\,(\rep{1},\rep{3})_0$ & $0$ & $+$\\
      \hline\str
    ~~$\mb{O}$ & ~~$\,(\rep{8},\rep{1})_0$ & $0$ & $+$\\
      \hline\str
    ~~$\mb{S}$ & ~~$\,(\rep{1},\rep{1})_0$ & $0$ & $+$\\
    \hline\str
    ~~$\mb{W^{\prime}}$ & ~~$\,(\rep{1},\rep{1})_0$ & $0$ & $+$ \\
    \hline
\end{tabular}
\caption{extended superpartners (adjoint chiral superfields). (Matter parity assignments shown for Sect.~\ref{sec4}.)}
\label{Tab2}
\end{center}
\end{table}
We observe that the model gives a sharp prediction of the Higgs boson mass of $141\pm2$ GeV. Removing the large uncertainty due to the unknown value of $\tan\beta$ results in a more precise Higgs mass prediction compared to Split SUSY or E-SSM. Notably, the result is relatively insensitive to the SUSY scale and order of magnitude changes to the SUSY scale $\widetilde{m}$ lead to only small ($\sim100\, \MeV$) deviations in the Higgs mass.

However, ATLAS and CMS searches  \cite{higgs} have recently confirmed the existence of a SM-like Higgs near $125$ GeV and this motivates the study of extensions of this minimal model which can give the correct Higgs mass.\footnote{Note, early LHC  searches \cite{LHC}  suggested a possible Higgs signal around 143 GeV, which partly motivated the minimal model \cite{Unwin:2011ag}.} One approach would be to introduce additional states at an intermediate scale to alter the RGE evolution (see e.g.~\cite{Liu:2012qua}).  In Sect.~\ref{sec4} we explore an alternative scenario in which the minimal spectrum is supplemented with adjoint chiral superfields which, as we shall discuss in the next section, can have a significant effect on the Higgs mass.

\section{High scale SUSY \& Dirac gauginos}
\label{sec3}

One of the most intriguing observations to arise from the recent Higgs discovery \cite{higgs} is that if one considers just the SM, then under renormalisation group evolution the Higgs quartic coupling appears to vanish in the UV \cite{van}. If we take this scenario seriously and treat it as a hint of the high energy theory in a similar vein to gauge coupling unification, then we would like a mechanism which sets $\lambda(\widetilde{m})\simeq0$  in the context of high scale SUSY. Whilst fixing $\tan\beta\simeq1$ will result in the quartic coupling vanishing at the SUSY scale, Hall \& Nomura have previously argued \cite{Hall:2009nd} that the scenario $\tan\beta\simeq1$ is {\em statistically disfavoured} compared to the large $\tan\beta$ scenario. However, there is a motivated way in which to set $\lambda(\widetilde{m})\simeq0$ in models of high scale SUSY independent of the value of $\tan\beta$.

If the MSSM spectrum is supplemented with extended superpartners (ESPs) -- chiral superfields which can provide Dirac mass terms for the gauginos as detailed in Tab.~\ref{Tab2} -- and the electroweak gaugino masses are mainly due to these Dirac mass terms, then the quartic coupling will approach zero at the SUSY scale. Moreover, these new fields are well motivated since such adjoint chiral superfields will enhance the SUSY of the gauge sector to $\mathcal{N}=2$. This scenario can be understood in the context of extra-dimensional models in which the gauge fields reside in the bulk while chiral matter fields are restricted to a 3-brane which only preserves $\mathcal{N}=1$ SUSY \cite{Fox:2002bu}.

%%%%%%%%%%%%%%%%%%%%%%%%%%%%%%%%%%%%
\begin{table}[t]
\begin{center}
\def\str{\vrule height12pt width0pt depth7pt}
\begin{tabular}{| c | l | c | c | }
    \hline\str
    ~Field ~&~ Gauge rep. ~&~ U(1)${}_R$ ~
    \\
    \hline\str
    ~~$\mb{H_u}$ & ~~$\,(\rep{1},\rep{2})_{1/2}$ & $0$ 
    \\
    \hline\str
    ~~$\mb{H_d}$ & ~~$\,(\rep{1},\rep{2})_{-1/2}$ & $0$ 
    \\
      \hline\str
    ~~$\mb{X}^{\prime}$ & ~~$\,(\rep{1},\rep{1})_0$ & $0$ 
    \\
    \hline
\end{tabular}
\vspace{5mm}
\caption{Replacing $\mb{H}$, $\mb{\eta}$, $\mb{X}$, in Tab.~\ref{Tab1} with  $\mb{H_u}$,  $\mb{H_d}$, $\mb{X}^{\prime}$ reintroduces the second Higgs doublet.}
\label{Tab3}
\end{center}
\end{table}
%%%%%%%%%%%%%%%%%%%%%%%%%%%%%%%%%%%%

Let us now return to the orthodox high scale SUSY framework with two Higgs doublets, this corresponds to a replacement of the fields  $\mb{H},~\mb{\eta}$ and $\mb{X}$ appearing in Tab.~\ref{Tab1} with $\mb{H_u},~\mb{H_d}$ and $\mb{X}^\prime$ from Tab.~\ref{Tab3}. We shall not initially insist that $R$-parity is conserved in this model. In this scenario it is assumed that only one linear combination of the Higgs states $H=H_u\sin\beta+H_d^{\dagger}\cos\beta$ is tuned light. The Higgs scalar mass matrix is of the form
\begin{equation}
\left(H_u^{\dagger},~H_d\right)
\left(\begin{array}{cc}
\widetilde{m}_{u}^2 & \widetilde{m}^2 \\
 \widetilde{m}^2 & \widetilde{m}_{d}^2
\end{array}
\right)
\left(\begin{array}{c}
H_u\\
H_d^{\dagger}
\end{array}\right)~,
\end{equation}
where $\widetilde{m}_{u,d}$ are the soft masses for $H_{u,d}$. Environmental selection of the weak scale \cite{hier} then requires that one of the eigenvalues of this matrix be of order $v$.

We now suppose that the fields are charged under a U(1) $R$-symmetry (as defined in Tab.~\ref{Tab1} - \ref{Tab3}) which, similar to Sect.~\ref{sec2}, forbids Majorana mass terms for the gauginos and ESPs and allows a $\mu$ term, and we assume that explicit Majorana mass terms for the ESPs are absent. Then comparable Majorana masses for the gauginos and ESPs arise from supergravity effects as in Sect.~\ref{sec2}, however, as these are parametrically the size of the gravitino mass $m_{3/2}\ll \widetilde{m}$, they will generically be smaller than the contributions from the Dirac mass terms which we expect to be at the SUSY scale. To be specific the new adjoint chiral superfields $\mb{O}$ and $\mb{T}$ and the singlet field  $\mb{S}$, allow the construction of the following Dirac mass terms for the gauginos \cite{Fox:2002bu,Dirac,Abel:2011dc,Benakli:2010gi} 
\begin{equation}
\begin{aligned}
\mathcal{L}_D=\int\dd^2\theta\frac{\mb{W^{\prime}_\alpha}}{M_*}\Big(&\lambda_3 \mathrm{Tr}(\mb{OW_3^{\alpha}})\\
&+\lambda_2 \mathrm{Tr}(\mb{TW_2^{\alpha}})+\lambda_1\mb{SW_1^{\alpha}}\Big)~,
\label{Dirac}
\end{aligned}
\end{equation}
where $\mb{W_i}$ are the gauge superfields of the SM gauge groups and $\mb{W^{\prime}}$ is a spurion vector superfield.  This leads to gaugino mass terms of the form
\begin{equation}
m_3 \mathrm{Tr}(\widetilde{O}\widetilde{G}) + m_2 \mathrm{Tr}(\widetilde{T}\widetilde{W}) + m_1 \mathrm{Tr}(\widetilde{S}\widetilde{B})~,
\label{Dirac2}
\end{equation}
where
\begin{equation}
m_i\simeq\frac{\lambda_i D^{\prime}}{M_*}~,
\label{Mi}
\end{equation}
and $D^\prime$ is the $D$-term SUSY breaking expectation value of a vector spurion superfield.

Importantly, the presence of the electroweak ESPs can result in a sizeable decrease in the Higgs mass, as they alter the SUSY boundary condition as follows \cite{Fox:2002bu}
 \begin{equation}
\lambda_H\simeq\frac{1}{8}\left(
\frac{M_1^2 g^\prime{}^2}{M_1^2 +4m_1^2}
+
\frac{M_2^2 g^2}{M_2^2 +4m_2^2}
\right)\cos^22\beta~,
\label{D}
\end{equation}
where $m_i$ is the Dirac mass for the bino/wino and $M_i$ is the Majorana mass of the associated ESP. The limit $M_i\ll m_i$ results in a $D$-flat direction and consequently the Higgs quartic coupling vanishes in this case. 
We can re-express this change in the boundary condition as a contribution to $\delta$, appearing in eq.~(\ref{ssbc}), of the form
 \begin{equation}
 \begin{aligned}
\delta_D\simeq&-\cos^2\theta_w\left(1-
\frac{M_2^2 }{M_2^2 +4m_2^2}\right)\\
&-\sin^2\theta_w\left(1-
\frac{M_1^2 }{M_1^2 +4m_1^2}\right)~.
\end{aligned}
\end{equation}
Moreover, for $M_i\ll m_i$ we can neglect the $\tan\beta$ dependence as this is subdominant compared to $\delta_D$.
We examine the Higgs mass calculation given in \cite{Hall:2009nd}, with $\delta\supset\delta_D$ and we find that  in the limit $m_1,m_2\rightarrow\infty$ the Higgs mass is given by 
\begin{equation}
\begin{aligned}
m_H\simeq&~127\,\GeV \\
&+1.3\, \GeV \left(\frac{m_t-173.1 \,\GeV}{0.9\, \GeV}\right)\\
&-0.35\, \GeV \left(\frac{\alpha_s(M_Z)-0.1176}{0.0007}\right)\\
&+0.14\, \GeV \log_{10}\left(\frac{\widetilde{m}}{10^{11} \,\GeV}\right)~.
\end{aligned}
\end{equation}
We compare this to the NLO calculation of the Higgs mass for high scale SUSY with $\lambda_H=0$ (i.e.~$\tan\beta=1$) given in \cite{Giudice:2011cg}, which found $m_H\simeq126\,\GeV$ for $\widetilde{m}\sim10^{11}$ GeV, in fair agreement. Note, \cite{Giudice:2011cg} also found that the scenario with $\lambda_H\simeq0$ is more sensitive to the value of the SUSY scale $\widetilde{m}$ than estimated above and the Higgs mass varies by around $0.5$ GeV per decade for $\widetilde{m}\gtrsim10^{10}$ GeV.
Furthermore, as the errors on the top mass and QCD coupling shrink, the dependence on $\widetilde{m}$ will become more prominent and limits on the scale $\widetilde{m}$ might be obtained from precision measurements of the Higgs mass. However, there is currently an irremovable theoretical error  $\sim 0.1$ GeV due to non-perturbative effects \cite{renormalon}.

In this scenario the QCD axion can play the role of the DM and anthropic arguments for axion DM relic abundance have been discussed at length in the literature \cite{axion}. Alternatively, if $R$-parity is conserved then the gravitino LSP could be the DM, however this scenario requires a dedicated study. Note that even if the gravitino LSP is stable due to $R$-parity, or some other stabilising symmetry, provided the reheat temperature is sufficiently low it need not present a significant contribution to the relic density. In the next section we outline a different scenario  in which matter parity is preserved and there is a WIMP DM candidate in the form of the neutral wino, similar to Sect.~\ref{sec2}.

%%%%%%%%%%%%%%%%%%%%%%%%

\section{E-SOHDM with Dirac Gauginos}
\label{sec4}

%%%%%%%%%%%%%%%%%%%%
\begin{figure*}[t]
\includegraphics[width=0.43\textwidth]{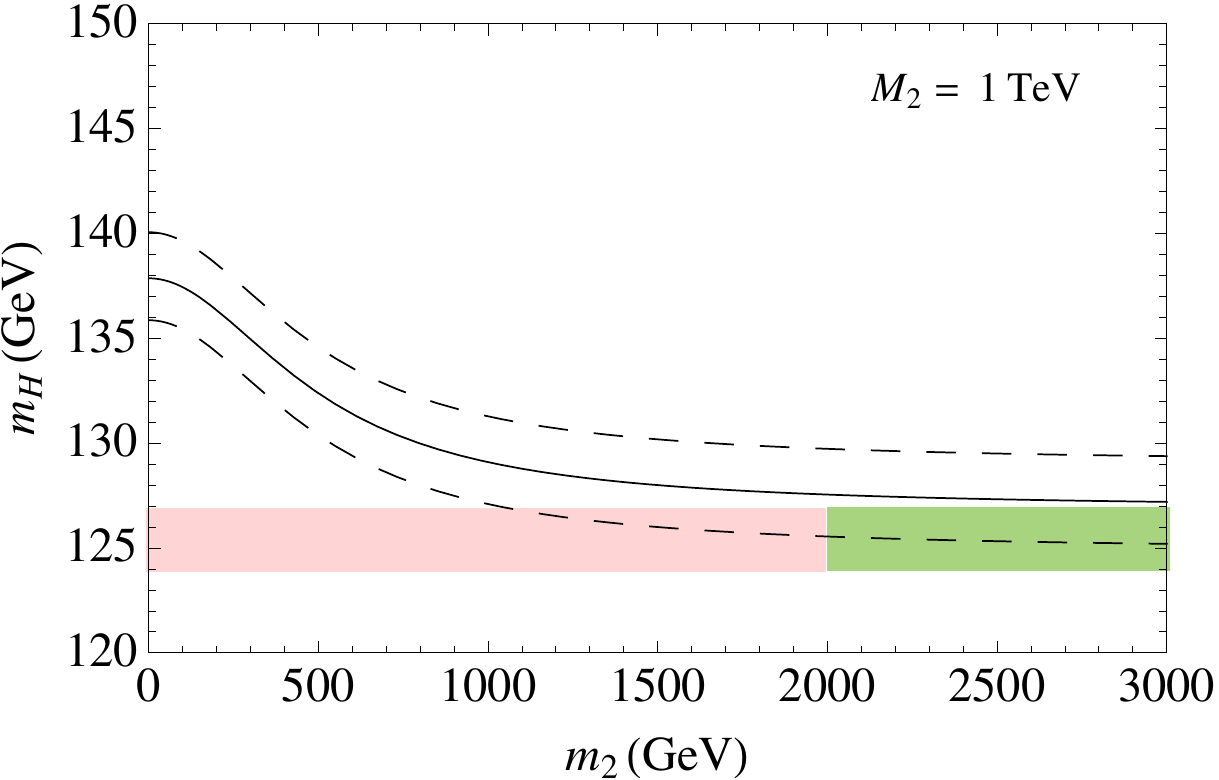} 
\hspace{0.5cm}
\includegraphics[width=0.43\textwidth]{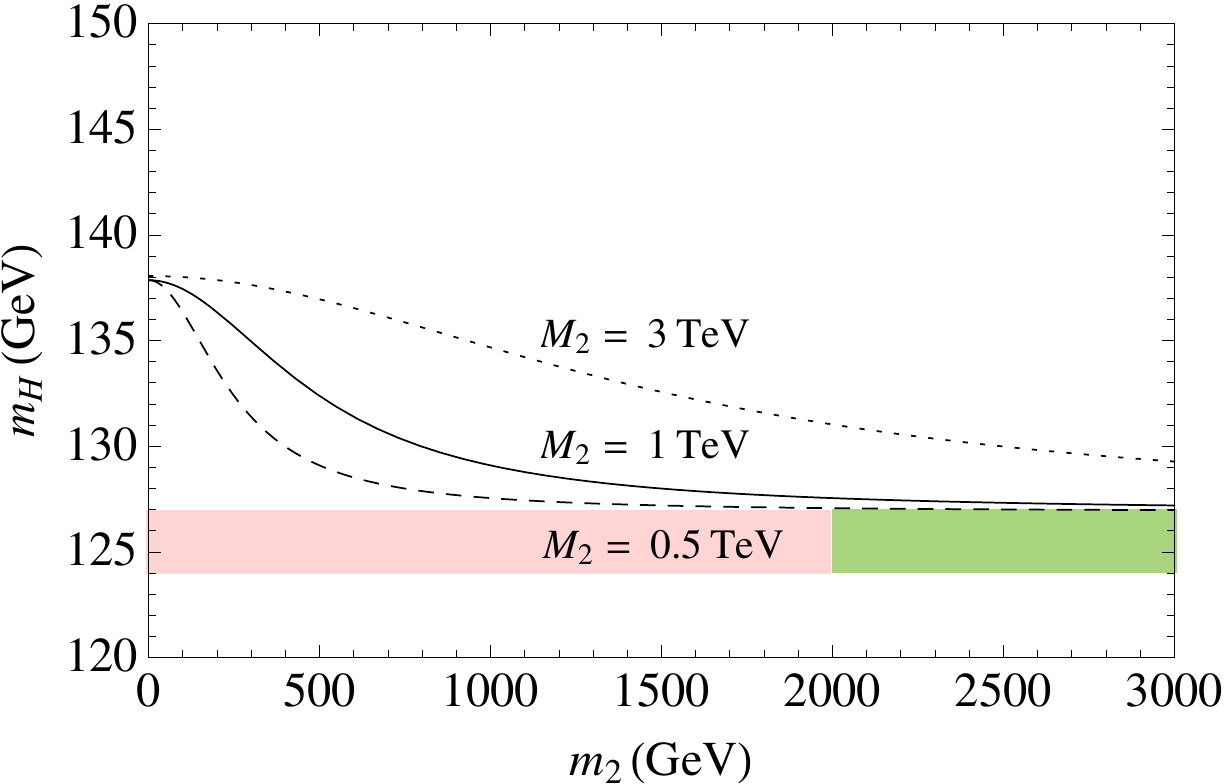}
\caption{
{\em Left.}~Higgs mass $m_H$ plotted against the wino Dirac mass $m_2$.  The red (light) shaded region indicates the mass range of the LHC Higgs discovery and the green (darker) shading shows the region in which the model also gives the correct DM relic density. We plot the case $M_2=1$ TeV, the long dashed upper (lower) line displays the effect of increasing (decreasing) the top mass by $0.9$~GeV (cf.~eq.~(\ref{mt})). We have assumed $\widetilde{m}=10^{11}$ GeV and taken $\alpha_S(M_Z)=0.1184$.
{\em Right.}~Similar to the first plot, but here we assume the central value for $m_t$ and display the cases $M_2=500$ GeV (dashed), 1 TeV (solid) and 3 TeV (dotted).
}
\label{Fig3}
\end{figure*}
%%%%%%%%%%%%%%%%%%%%%

\vspace{-2mm}

We next construct a model which combines the strengths of the two scenarios explored previously, thus we gain the aesthetic appeal of only a single Higgs doublet, WIMP DM, and a Higgs mass which agrees with the observed value. We consider the spectrum given in Tab.~\ref{Tab1},  supplemented with the extended superpartners of Tab.~\ref{Tab2}. 
With this extended spectrum, Dirac mass terms can be constructed for the gauginos as detailed in eq.~(\ref{Dirac}) \& (\ref{Dirac2}). Let us suppose that the low energy spectrum of the model contains only the SM states, the winos and the corresponding scalar ESP states. The neutralino can be the LSP, as the associated scalar adjoints are generically heavier \cite{Benakli:2010gi} and we shall focus on the case of a wino LSP. To obtain a suitable splitting in the spectrum we shall assume that   
$D^{\prime}, F_X\sim\widetilde{m}M_*$
 and that $\lambda_2$ (alternatively $\lambda_1$) is tuned small through environmental selection on the mass of the wino (bino) DM. Models with comparable  $D$- and $F$- term breaking have been studied in \cite{SB}.  With natural couplings $\lambda_1,\, \lambda_3\sim 1$ the Dirac bino and gluino, and the associated scalar ESPs have masses $\sim\widetilde{m}$.  Since in this scenario the wino Dirac mass is environmentally selected we take $M_2<m_2$. Indeed it is permissible that $M_2\ll 1$ TeV in which case, since $M_i\sim m_{3/2}$, the scale of SUSY breaking could be much lower. 

The annihilation cross section of Majorana neutralinos is p-wave suppressed, but this is not the case for Dirac states. Whilst this has a large effect on Dirac bino DM \cite{bino}, the relic density for Dirac wino DM, which is mainly set by the coannihilation rate, is relatively unaffected and the correct relic density will still be obtained for $\sim3$ TeV. Combining this requirement with eq.~(\ref{Mi}) gives
\begin{equation}
 \widetilde{m}\sim \frac{3\,\TeV}{\lambda_2}~. 
\end{equation}
Comparing this stipulation with eq.~(\ref{con}), which ensures suitable SM Yukawa couplings for models with one Higgs doublet, leads to the  condition:
\begin{equation}
M_*\gtrsim \frac{100 \,\TeV}{\lambda_2 }~.
\end{equation}
Note, the presence of TeV winos requires that $\lambda_2\ll 1$ and thus $M_*$ can be sufficiently large to  generate appropriate neutrino masses via the Weinberg operator as in Sect.~\ref{sec2}. 
  
 Given $\widetilde{m}<M_*$, new contributions from $\mb{S}$ and $\mb{T}$ will only lead to small deviations in the RGE evolution,\footnote{Couplings of the ESP fields to the Higgs are via
$
\int\dd^4\theta\frac{\mb{X^{\dagger}}}{M_*^2}\Big(\lambda_S \mb{SH\eta} + \lambda_T \mb{HT\eta}\Big).
$
The operators $\mb{HS\eta}$ and  $\mb{HT\eta}$ contribute to the running of the Higgs quartic coupling at the order
$
\mathcal{A}_{S,T}\sim\frac{1}{16\pi^2}\left(\frac{\lambda_{S,T} F_X}{M_*^2}\right)^4
\sim \frac{\lambda_{S,T}^4 \widetilde{m}^4}{16\pi^2 M_*^4},
$
and thus are suppressed for $\widetilde{m}< M_*$.}
more significantly, the Dirac mass terms alter the SUSY boundary condition\footnote{This corrects discussions in \cite{Unwin:2011ag} where this effect was neglected.} as given in eq.~(\ref{D}) and since only the winos remain light and the bino and gluino have Dirac masses at $\sim\widetilde{m}$, the correction $\delta_D\subset\delta$ is as follows
 \begin{equation}
 \begin{aligned}
\delta_D\simeq&-\left(1
-\cos^2\theta_w\frac{M_2^2 }{M_2^2 +4m_2^2}
\right)~.
\end{aligned}
\end{equation}
In Fig.~\ref{Fig3}, left panel, we plot the dependence of the Higgs mass on the magnitude of the Dirac mass of the wino for $M_2=1~\TeV$, including the uncertainties due to $m_t$ (which lead to $\sim\pm2$ GeV error in the Higgs mass). In the right panel of Fig.~\ref{Fig3} we consider how the Higgs mass dependence on $m_2$ changes as the Majorana mass $M_2$ is varied. 
We assume that environmental selection on the Dirac mass $m_2$  results in the correct wino mass and we find that for $M_2\lesssim1$ TeV the Higgs mass is required to be $127\pm2$ GeV (this value rises slightly for $1~\TeV\lesssim M_2\leq m_2$). This result is consistent with the ATLAS and CMS measurements of the Higgs mass \cite{higgs}, which are, respectively, $(126~\pm~0.4~\pm~0.4)$ GeV and $(125.3~\pm~0.4~\pm~0.5)$ GeV (with statistical and systematic errors).  As in the previous models, we obtain a sharp result, because the $\tan\beta$ dependence is removed. The main source of uncertainty is still from $m_t$ and, similarly, there is only a weak dependence on $\widetilde{m}$. Thus, in order to obtain a Higgs mass consistent with LHC measurements we require that the winos be dominantly Dirac and consequently the quartic Higgs coupling will be small at the the SUSY scale:
$\lambda_H(\widetilde{m})\lesssim0.1$.

Before closing this section we note that the ESP fields can be embedded into an adjoint representation of $SU(5)$ \cite{Benakli:2010gi}. 
In order to complete the adjoint representation we introduce a pair of vector-like `bachelor' superfields $(\rep{3},\rep{2})_{-5/6}$ and $(\rep{3},\overline{\rep{2}})_{ 5/6}$ with masses at the scale $\widetilde{m}$. 
There is a danger that the hypercharge ESP singlet field $\mb{S}$ may acquire a large tadpole term, however this can be avoided if the couplings to the messengers respect $SU(5)$ \cite{Abel:2011dc} or are otherwise suitably arranged \cite{Benakli:2010gi}.

%%%%%%%%%%%%%%%%%%%%%%%%

\section{Distinguishing Models}
\label{sec5}

We shall make some brief comments on how models with high scale SUSY breaking could be probed and different proposals might be distinguished. We will focus on the models of Sect.~\ref{sec3} \& \ref{sec4}, in which the gauginos are (pseudo)-Dirac and the correct Higgs mass can be obtained. In particular, the U(1) $R$-symmetry can lead to distinctive collider signatures which could be used to differentiate these models from other theories of high scale SUSY breaking with $\tan\beta\simeq1$.

Note that in models with (pseudo)-Dirac gauginos the available production and decay channels of the gauginos and sfermions are altered. Whilst most of these states lie beyond the reach of current technology, indirect searches may be possible. An analysis of the ratio of like to unlike sign di-lepton events with large missing energy could potentially determine the nature of the gauginos if they are near the weak scale \cite{Choi:2008pi}. As the sfermions are expected to be very heavy this analysis would require a large amount of data and careful study. If the ESP fields are light then these could also lead to distinct signals which could be used to distinguish these models from alternative proposals, see for example~\cite{ESP}.

Whilst it is not inconceivable that the effects of the TeV scale gauginos could be detected in a next-generation collider, perhaps a more immediately accessible window on models with environmental selection is provided by observations of the early universe. In particular, deviations during Big Bang nucleosynthesis (BBN). In the E-SOHDM framework the bino has no direct decay route and must first mix with the Higgsino, while the gluinos can only decay via heavy squarks. Consequently, both binos and gluinos potentially have long lifetimes and their late decays could result in observable signals during BBN. The effects of decaying gluinos during BBN have been consider in the context of Split SUSY \cite{split} and general constraints on energy injection during BBN are studied in \cite{BBN}. These cosmological constraints can be ameliorated if one assumes that the reheating temperature is less than the gluino/bino mass, such that these states can not be produced after reheating.

Also note that to allow the cosmological constant to be adjusted sufficiently close to zero, the $R$-symmetry must be broken at high scales by supergravity effects and this results in an $R$-axion. Since the SUSY breaking scale is high the $R$-axion is heavy \cite{Raxion}
\begin{equation}
m_{R_a}^2\sim\frac{|F_X|^{3/2}}{M_{\mathrm{Pl}}}\sim \left(10^{9}\,\GeV\right)^2\left(\frac{\sqrt{F}}{10^{12}~\GeV}\right)^{3}~.
\label{mar}
\end{equation}
Consequently, for high scale SUSY breaking the  $R$-axion, and likewise the gravitino ($m_{3/2}\sim500$ TeV), are heavy enough to evade all cosmological constraints and searches. Whilst the scale of the SUSY breaking is not sufficiently high in order to avoid all cosmological problems due to moduli (specifically, the modulus field associated to the overall volume has a mass $\sim1$ GeV), discussions on circumventing these difficulties can be found in  \cite{Conlon:2007gk}. However, as noted in Sect.~\ref{sec4}, if environmental selection acts on the Dirac mass of the wino to provide the DM relic density, then the scale of SUSY breaking can be lower, and this could potentially result in cosmological signals from the gravitino, $R$-axion or moduli \cite{Raxion,Conlon:2007gk}.

%%%%%%%%%%%%%%%%%%%%%%%%

\section{Concluding remarks}

We have explored some of the model building possibilities available for models of high scale SUSY with an $R$-symmetry. In particular, we presented a new class of high scale SUSY models by applying the principles of environmental selection to the Supersymmetric One Higgs Doublet Model and discussed the important effects of Dirac gauginos on the SUSY boundary condition. Specifically, we demonstrated that models with Dirac gauginos and a SUSY scale at $10^{10\pm1}$ GeV naturally result in a Higgs mass near $125$ GeV, since the Higgs quartic coupling vanishes at the SUSY scale.

We discussed various manners in which the models presented here can be discerned from existing models. One of the main differences between models of high scale SUSY is the source of the DM relic density; axions, Majorana-winos, Dirac-binos/winos and mixed wino-Higgsino neutralinos (as in Split SUSY) lead to different predictions and thus may allow these competing models to be distinguished. Furthermore, in Sect.~\ref{sec4} we constructed a phenomenologically interesting model with one Higgs doublet, WIMP DM, and a Higgs mass of $127\pm2$ GeV and this scenario makes a number of testable predictions such as Dirac gauginos and WIMP DM with a mass around 3 TeV.

Although supersymmetry or new strong dynamics could ultimately resolve the hierarchy problem, if only the SM Higgs is found after the full LHC run with $100\,\fb^{-1}$ then more radical ideas must be seriously contemplated.  If the LHC discovers only a single Higgs and no signals of physics beyond the SM, then it becomes highly plausible that fine-tuning is inherent to the physical universe. The existence of such fine-tuning would lend exceptional credence to the concept of the  multiverse.

\section*{Acknowledgements} 

I am grateful to John March-Russell and Stephen West for useful discussions. I would also like to thank the referees for their helpful comments and  Matthew McCullough for comments on an early draft. This work was partially funded by an EPSRC doctoral studentship and awards from St. John's College \& Pembroke College, Oxford.

\end{document}